# Discreteness of curved space-time from GUP

**Ahmad Adel Abutaleb**

Dept of Mathematics, , Faculty of Science, University of Mansoura, Elmansoura 35516, Egypt

**Abstract**

Diverse theories of Quantum Gravity expect modifications of the Heisenberg uncertainty principle near the Planck scale to a so-called Generalized uncertainty principle (GUP). It was shown by some authors that the GUP gives rise to corrections to the Schrodinger , Klein-Gordon and Dirac equations. By solving the GUP corrected equations, the authors arrived at quantization not only of energy but also of box length, area and volume. In this paper, we extend the above results to the case of curved space–time (Schwarzschild metric). We showed that we arrived at the quantization of space by solving Dirac equation with GUP in this metric.

**Introduction**

Diverse approaches to quantum gravity expect a minimum measurable length, and a modification of the Heisenberg uncertainty principle to a so-called generalized uncertainty principle or GUP. This implies a modification of the commutation relations between position coordinates and momentum. In [1] the following proposed GUP consistent with Doubly special relativity or DSR theories and black hole physics which ensure $[x_i, x_j] = 0 = [p_i, p_j]$:

$$[x_i. p_j] = ih[\delta_{ij} - a(p\delta_{ij} + \frac{p_i p_j}{p}) + a^2(p^2 \delta_{ij} + 3p_i p_j)],$$

$$\Delta x \Delta p \geq \frac{h}{2}[1 - 2a<p> + 4a^2<p^2>]$$

$$\geq \frac{h}{2}[1 + \left(\frac{a}{\sqrt{p^2}} + 4a^2\right)\Delta p^2 + 4a^2<p>^2 - 2a\sqrt{<p^2>}], \tag{1}$$

where,
a=$a_0/M_{pl}c = a_0 L_{pl}/h$, $M_{pl} = planck\ mass$, $L_{pl} \simeq 10^{-35}m = planck\ length$, $M_{pl}c^2 = planck\ energy \approx 10^{19} GeV$. GUP induced terms become important near the Planck scale. ( for earlier version of GUP motivated by string theory, black hole physics, DSR, see [2-14], for some phenomenological implications see [1,15,16].

Equations (1) imply the following minimum measurable length and maximum measurable momentum [1,17]

$$\Delta x \geq (\Delta x)_{min} \approx a_0 L_{pl},$$

$$\Delta p \leq (\Delta p)_{max} \approx M_{pl}c/a_0. \tag{2}$$

It is natural to take $a_0 = 1$, for more details see [17].

The following definitions proposed in [1] and used in [1,17]

$$x_i = x_{0i} \, , \, p_i = p_{0i}(1 - ap_0 + 2a^2 p_0^2) \tag{3}$$

with $x_{0i}, p_{0j}$ satisfying the canonical commutation relations $[x_{0i}, p_{0j}] = ih\delta_{ij}$, such that $p_{0i} = -ih\frac{\partial}{\partial x_{0i}}, p_0^2 = \sum_{j=1}^{3} p_{0j} p_{0j}$.

In [1] it was shown that any non-relativistic Hamiltonian of the form $H = p^2/2m + V(\vec{r})$ can be written as $H = p_0^2/2m - (a/m)p_0^3 + V(\vec{r}) + O(a^2)$ using (3). This corrected Hamiltonian implies not only the usual quantization of energy, but also that the box length is quantized. In [17] the above results were extended to a relativistic particle in two and three dimensions. In this paper we study Dirac equation in Schwarzschild metric using GUP and show that we arrive at the quantization of space.

**GUP Dirac equations in Schwarzschild metric**

Dirac equation in Schwarzschild metric without (GUP) can be written as follow [18]

$$(c(\vec{\alpha}.\vec{p}) + \beta mc^2)\psi = \frac{E}{\sqrt{\zeta}}\psi \tag{4}$$

where $m$ is the rest mass of the particle, $\psi$ is the Dirac spinor, $\vec{\alpha}$ and $\beta$ are Dirac matrices, $\vec{p} \equiv (p_r, p_\theta, p_\varphi)$ are momentum operators, $\zeta = 1 - r_s/r$, $r_s$ is the Schwarzschild radius of massive body, related to its mass $M$ by $r_s = 2GM/C^2$, $G$ is the gravitational constant, $c$ is the speed of light in free space. using $\psi = \begin{pmatrix} \chi_1 \\ \chi_2 \end{pmatrix}$, (4) can be written as

$$c(\vec{\alpha}.\vec{p})\chi_2 + mc^2 \chi_1 = \frac{E}{\sqrt{\zeta}}\chi_1 \, ,$$

$$c(\vec{\alpha}.\vec{p})\chi_1 - mc^2 \chi_2 = \frac{E}{\sqrt{\zeta}}\chi_2 \, . \tag{5}$$

Now, using GUP correction (3) and (5) take the form

$$c(\vec{\alpha}.\vec{p_0})\chi_2 + mc^2 \chi_1 - cap_0^2 \chi_1 = \frac{E}{\sqrt{\zeta}}\chi_1 \, ,$$

$$c(\vec{\alpha}.\vec{p_0})\chi_1 - mc^2 \chi_2 - cap_0^2 \chi_2 = \frac{E}{\sqrt{\zeta}}\chi_2 \, . \tag{6}$$

where, $p_0^2 = -h^2 [\frac{\sqrt{\zeta}}{r^2}\frac{\partial}{\partial r}(r^2 \sqrt{\zeta}\frac{\partial}{\partial r}) + \frac{1}{r^2 \sin\theta}\frac{\partial}{\partial \theta}(\sin\theta \frac{\partial}{\partial \theta}) + \frac{1}{r^2 \sin^2\theta}\frac{\partial^2}{\partial \varphi^2}]$. \hfill (7)

We study Dirac equations in (6) in Schwarzschild metric with spherical cavity with radius R defined by the potential

$$U(r) = 0 \quad , r \leq R \, ,$$

$$U(r) = U_0 \to \infty \quad , r > R \tag{8}$$

so, we can write the corrected GUP Dirac equations with spherical cavity defined by (8) in Schwarzschild metric as

$$c(\vec{\alpha}.\overrightarrow{p_0})\chi_2 + (mc^2 + U)\chi_1 - ca p_0^2 \chi_1 = \frac{E}{\sqrt{\zeta}}\chi_1$$

$$c(\vec{\alpha}.\overrightarrow{p_0})\chi_1 - (mc^2 + U)\chi_2 - ca p_0^2 \chi_2 = \frac{E}{\sqrt{\zeta}}\chi_2 \,. \tag{9}$$

Notice that, when $a = 0, \zeta = 1$, equations in (9) are usual Dirac equations in flat space-time. When $a \neq 0, \zeta = 1$, equations (9) are Dirac equations with GUP in flat space-time proposed in [19]. When $a = 0, \zeta \neq 1$, equations in (9) are Dirac equations in Schwarzschild metric without GUP defined in [18]. We follow the analysis of [17,19] and related references [20,21].

We assume the form of Dirac spinor as

$$\psi = \begin{pmatrix} \chi_1 \\ \chi_2 \end{pmatrix} = \begin{pmatrix} g_k(r)\gamma_{jl}^{j_3}(\hat{r}) \\ if_k(r)\gamma_{jl^-}^{j_3}(\hat{r}) \end{pmatrix} \tag{10}$$

$$\gamma_{jl}^{j_3}(\hat{r}) = \left(l \; \frac{1}{2} \; j_3 - \frac{1}{2} \; \frac{1}{2} \middle| j \; j_3\right) Y_l^{j_3-\frac{1}{2}}(\hat{r})\begin{pmatrix}1\\0\end{pmatrix}$$

$$+ \left(l \; \frac{1}{2} \; j_3 + \frac{1}{2} \; \frac{-1}{2} \middle| j \; j_3\right) Y_l^{j_3+\frac{1}{2}}(\hat{r})\begin{pmatrix}0\\1\end{pmatrix} \tag{11}$$

where, $Y_l^{j_3-\frac{1}{2}}(\hat{r})$ and $Y_l^{j_3+\frac{1}{2}}(\hat{r})$ are spherical harmonics and

$(j_1 \; j_2 \; m_1 \; m_2 | j \; j_3)$ are Clebsh-Gordon coefficients, $\chi_1, \chi_2$ are eigenstates of $L^2$ ($\vec{L}$ is the angular momentum operator) with eigenvalues $h^2 l(l+1)$ and $h^2 l^-(l^-+1)$ respectively, such that the following hold

If $k = j + \frac{1}{2} > 0$ \hfill (12)

then $l = k = j + \frac{1}{2} \;,\; l^- = k - 1 = j - \frac{1}{2}$ \hfill (13)

and if $k = -(j + \frac{1}{2}) < 0$ \hfill (14)

then $l = -(k+1) = j - \frac{1}{2} \;,\; l^- = -k = j + \frac{1}{2}$ . \hfill (15)

We use $(\vec{\alpha}.\vec{A})(\vec{\alpha}.\vec{B}) = \vec{A}.\vec{B} + i\vec{\alpha}.(\vec{A} \times \vec{B})$ and the related identity $(\vec{\alpha}.\vec{r})(\vec{\alpha}.\vec{r}) = r^2$, so we have

$$(\vec{\alpha}.\overrightarrow{p_0}) = \frac{(\vec{\alpha}.\vec{r})(\vec{\alpha}.\vec{r})(\vec{\alpha}.\overrightarrow{p_0})}{r^2} = \frac{\vec{\alpha}.\vec{r}}{r^2}\left(\vec{r}.\overrightarrow{p_0} + i\vec{\alpha}.(\vec{r}\times\overrightarrow{p_0})\right) = \frac{\vec{\alpha}.\vec{r}}{r^2}\left(\vec{r}.\overrightarrow{p_0} + i\vec{\alpha}.\vec{L}\right). \tag{16}$$

But from the definition of the momentum operators in Schwarzschild metric [18], we can write (16) as

$$(\vec{\alpha}.\overrightarrow{p_0}) = (\vec{\alpha}.\hat{r})\left(-ih\sqrt{\zeta}\frac{\partial}{\partial r} + \frac{i}{r}\vec{\alpha}.\vec{L}\right). \tag{17}$$

Also, we have

$$(\vec{\alpha}.\vec{L}+1)\chi_{1,2} = \mp\chi_{1,2}.$$

$$(\vec{\alpha}.\hat{r})\,\gamma_{jl}^{j_3}(\hat{r}) = -\gamma_{jl^-}^{j_3}(\hat{r})\quad,\quad (\vec{\alpha}.\hat{r})\gamma_{jl^-}^{j_3}(\hat{r}) = -\gamma_{jl}^{j_3}(\hat{r}).\tag{18}$$

Next, from the definition of $p_0^2$ in Schwarzschild metric [18] we can write

$$p_0^2 F(r)Y_l^m = h^2\left[-\frac{\sqrt{\zeta}}{r^2}\frac{d}{dr}\left(r^2\sqrt{\zeta}\frac{\partial}{\partial r}\right) + \frac{l(l+1)}{r^2}\right]F(r)Y_l^m\tag{19}$$

so, using equations (17),(18), and(19), we can obtain from (9) the following equations:

$$-ch\sqrt{\zeta}\frac{df_k}{dr} + \frac{c(k-1)}{r}f_k + (mc^2+U)g_k + cah^2\left[\frac{\sqrt{\zeta}}{r^2}\frac{d}{dr}\left(\sqrt{\zeta}r^2\frac{dg_k}{dr}\right) - \frac{l(l+1)}{r^2}g_k\right] = \frac{E}{\sqrt{\zeta}}g_k$$

$$ch\sqrt{\zeta}\frac{dg_k}{dr} + \frac{c(k+1)}{r}g_k - (mc^2+U)f_k + cah^2\left[\frac{\sqrt{\zeta}}{r^2}\frac{d}{dr}\left(\sqrt{\zeta}r^2\frac{df_k}{dr}\right) - \frac{l^-(l^-+1)}{r^2}f_k\right] = \frac{E}{\sqrt{\zeta}}f_k.\tag{20}$$

It can be shown that MIT bag boundary condition (at $r = R$) is equivalent to [19,20]

$$\tilde{\psi}\psi = 0.\tag{21}$$

As in [17], we can expect new nonperturbative solutions of the form $f_k = F_K(r)e^{i\epsilon r/ah}$, $g_k = G_K(r)e^{i\epsilon r/ah}$ (where $\epsilon = O(1)$) for which equations (20) simplifies to

$$ah\frac{d^2g_k}{dr^2} = \sqrt{\zeta}\frac{df_k}{dr}$$

$$ah\frac{d^2f_k}{dr^2} = -\sqrt{\zeta}\frac{dg_k}{dr}\tag{22}$$

where we have dropped terms which are ignorable for small $a$.

When $\zeta = 1$, equations in (22) are identical to the equations (60) - (61) in reference [17], and in this case we have the following solutions $f_k^N = iNe^{ir/ah}$, $g_k^N = Ne^{ir/ah}$, $N$ is constant, so we can assume the solutions of (22) as

$$f_k^N = iNe^{ir/ah}$$

$$g_k^N = NB(r)e^{ir/ah}.\tag{23}$$

By applying equations (23) on (22), we find that

$$a^2h^2\frac{d^2B(r)}{dr^2} + B(r) + \sqrt{\zeta} - \frac{2}{\sqrt{\zeta}} = 0.\tag{24}$$

Consider that $r_s$ is very small, so we can approximate equation (24) to

$$a^2h^2\frac{d^2B(r)}{dr^2} + B(r) - 1 - \frac{3r_s}{2r} = 0.\tag{25}$$

The solution of (25) take the form

$$B(r) = c_1\sin(r/ah) + c_2\cos(r/ah) + \frac{1.5r_s}{ah}[C_i(r/ah)\sin(r/ah) - S_i(r/ah)\cos(r/ah)] + 1 \quad (26)$$

where, $c_1$ and $c_2$ are constants, $S_i(r/ah)$ and $C_i(r/ah)$ are the sine integral function and cosine integral function defined as

$$S_i(y) = \int_0^y \frac{SINt}{t}dt \quad \text{and} \quad C_i(y) = -\int_y^\infty \frac{COSt}{t}dt$$

. For more details about this functions see [22,23].

Therefore, the solutions of (22) take the form

$$f_k^N = iNe^{ir/ah},$$

$$g_k^N = N[1 + \sigma(r)]e^{ir/ah} \quad (27)$$

where,

$$\sigma(r) = c_1\sin(r/ah) + c_2\cos(r/ah)$$

$$+ \frac{1.5r_s}{ah}[C_i(r/ah)\sin(r/ah) - S_i(r/ah)\cos(r/ah)]. \quad (28)$$

Here, one must have $\lim_{r_s \to o} c_1 = \lim_{r_s \to o} c_2 = 0$, and in this case ($\sigma(r) = 0, r_s = 0$), the results are the same of [17].

Now, the boundary condition (21) gives

$$|g_k(R) + g_k^N(R)|^2 = |f_k(R) + f_k^N(R)|^2 \quad (29)$$

which to $O(a)$ translate to

$$(g_k^2 - f_k^2) + 2N[g_k(1 + \sigma(R))\cos(R/ah) - f_k\sin(R/ah)] + N^2[(1 + \sigma(R))^2 - 1] = 0 \quad (30).$$

From (30), we have

$$f_k = g_k. \quad (31)$$

$$\tan(R/ah) = 1 + \sigma(R). \quad (32)$$

$$\sigma(R) = 0, or\ \sigma(R) = -2.$$

Observing equation (28), we can write

$$\frac{\sigma(R)}{\cos(R/ah)} = c_1\tan(R/ah) + c_2 + \frac{1.5r_s}{ah}[C_i(R/ah)\tan(R/ah) - S_i(R/ah)]. \quad (33)$$

From (32) we have

if $\sigma(R) = 0$ then $\tan(R/ah) = 1$ and if $\sigma(R) = -2$ then $\tan(R/ah) = -1$. (34)

For the case of $\sigma(R) = 0$, $r_s = 0$ we have the same result of discreteness of space in flat Space-Time [17].

For the second case $\sigma(R) = -2$, we have, from (33),

$$\frac{4ah}{3r_s COS(R/ah)} + \frac{2Cah}{3r_s} = C_i(R/ah) + S_i(R/ah) \quad , C \text{ is constant} \tag{35}$$

If we take the constant $C = 0$, then we have

$$\frac{4ah}{3r_s COS(R/ah)} = C_i(R/ah) + S_i(R/ah) . \tag{36}$$

Suppose that we choose $r_s = \frac{4}{3} ah$ (very small as assumption); then we have

$$\frac{1}{COS(R/ah)} = C_i(R/ah) + S_i(R/ah) . \tag{37}$$

Equation (37) has infinitely number of solutions, we write some numerical values of it, $(R/ah) = 5,565$, $(R/ah) = 7.159$, $(R/ah) = 11,755$, $(R/ah) = 13.448$ so, the radius of cavity $R$ has been quantized in terms of $ah$ and we again arrived at the quantization of space in Schwarzschild-like metric.

**Conclusion**

Dirac equations with GUP in Schwarzschild metric have been studied. We showed that the assumption of existence of a minimum measurable length and a corresponding modification of uncertainty principle yields discreteness of space in this metric. But the question now arises, what is the guarantee that this result will continue to hold for more generic curved spacetimes? We expect that this discreteness will always appear provided that generalized uncertainty principle enters into the theory, but in fact we have no mathematical proof of existence of such discreteness of space if we work on the general metric of the general relativity theory.

**References**

[1] A.F. Ali, S. Das, E.C. Vagenas, Phys.Lett.B, vol. **678** (2009) 497 [arxiv:0906.5396[hep-th]].

[2] D. Amatti, M. Ciafaloni, G. Veneziano, Phys.Lett.B, vol. **216** (1980) 41.

[3] M. Maggiore, Phys.Lett.B, vol. **304** (1993) 65 [arxiv:hep-th/9301076].

[4] M. Maggiore, Phys.Rev.D, vol. **49** (1994) 5182 [arxiv:hep-th/9305165].

[5] F. Scardigli, Phys.Lett.B, vol. **452** (1999) 39 [arxiv:hep-th/9904025].

[6] S. Hossenfelder, M. Bleicher, S. Hofmann, J. Ruppert, S. Scherer, H. Stoecker, Phys.Lett.B, vol. **575** (2003) 85 [arxiv:hep-th/0305262].

[7] A. Kempf, G. Mangano, R.B. Mann, Phys.Rev.D, vol. **52** (1995) 1108 [arxiv:hep-th/9412167].

[8] A. Kempf, J.Phys.A, vol. **30** (1997) 2093 [arxiv:hep-th/9604045].

[9] F. Brau, J.Phys.A, vol. **32** (1999) 7691 [arxiv:quant-ph/9905033].

[10] G.A. Camelia, Int.J.Mod.Phys.D, vol. **11** (2002) 35 [arxiv:gr-qc/0012051].

[11]  J. Magueijo, L. Smolin, Phys.Rev.D, vol. **67** (2003) 044017 [arxiv:gr-qc/0207085].

[12] J.L. Cortes, J. Gamboa, Phys.Rev.D, vol **71** (2005) 065015 [arxiv:hep-th/0405285].

[13] A. Ashoorioon, R.B. Mann, Nuclear Physics B **716** (2005) 261 [arxiv:gr-qc/0411056].

[14] A. Ashoorioon, A. Kempf, R.B. Mann, Phys Rev D, vol. **71** (2005) 023503 [arxiv:astro-ph/0410139].

[15] S. Das, E. Vagenas, Phys.Rev.Lett, vol. **101** (2008) 221301 [arxiv:0810.5333[hep-th]].

[16] S. Das, E. Vagenas, Can.J.Phys, vol. **87** (2009) 233 [arxiv:0901.1768[hep-th]].

[17] S. Das, E.C. Vagenas , A.F. Ali, Phys.Lett.B, vol. **690** (2010) 407 [arxiv:1005.3368[hep-th]].

[18] C.C. Barros Jr, Eur.Phys.J, vol. **42** (2005) 119 [arxiv:gen-ph/0409064].

[19] R.K. Bhadri, Models of the Nucleon, from Quarks to Solitions, Addison Wesley (1988), Chp 2.

[20] A.W. Thomas, Adv.Nucl.Phys, vol. **13** (1984)  1.

[21] A.J. Hey. Lecture notes in physics **77** (1978) 155 (Spriger-Verlag, Berlin).

[22] J. Havil, Gamma:Exploring Eulers constant, Princeton University Press (2003).

[23] M. Abramowits, I.A. Stegun, Handbook Of Mathematical Functions with Formulas , Graphes and Mathematical Tables, New York: Dover (1972) .